\documentstyle[preprint,aps,epsf,psfig]{revtex}
\begin{document}
\title{
       Prospects for Parity Non-conservation Experiments 
       with Highly Charged Heavy Ions
      }
\author{ 
        M.~Maul, A.~Sch{\"a}fer, W.~Greiner}
\address{
        Institut f{\"u}r Theoretische Physik, Johann Wolfgang Goethe-Universit{\"a}t,\\
        D-60054 Frankfurt am Main, Germany}

\author{
        P.~Indelicato}
\address{
	Laboratoire de Physique Atomique et Nucl\'eaire, \\
	Unit\'e de Recherche soci\'ee au CNRS no. 771, \\Universit\'e Pierre et
	Marie
	Curie,\\Case 93, 4
	place Jussieu, F-75252 Paris CEDEX 05, France }
\date{revised by P.I.  Jan 9 1996 }
\maketitle
\hfill UFTP preprint 408/1996 
\begin{abstract}
We discuss the prospects for parity non-conservation experiments with
highly charged heavy ions. Energy levels and parity mixing for heavy
ions with two to five electrons are calculated. We investigate
two-photon-transitions and the possibility to observe interference
effects between weak-matrix elements and Stark matrix elements for
periodic electric field configurations.
\end{abstract}
\pagebreak

\section{Introduction}
Atomic physics tests of the standard model \cite{1,2,3} play a very
special role because of the small momentum transfers
involved. Comparisons between their results and high energy data are
highly sensitive to radiative corrections and thus to extensions of
the standard model \cite{5}. With the percent precision reached in the
Cs experiments described in \cite{3}, the effect of radiative
corrections is of the order of the experimental accuracy. If a system
is found for which a 0.1\% accuracy can be reached the experimental
results would allow most interesting and far reaching conclusions (see
e.g. \cite{ros95}). For the atoms and experimental setups studied so
far this seems unluckily to be out of reach, which motivates the
search for significantly different alternatives. The possibility we
want to discuss is the use of highly charged heavy ions, which can be
produced and stored in great variety at, e.g., Gesellschaft f\"ur
Schwerionenforschung in Darmstadt, Germany. We discussed already some
time ago the prospects for inducing a two-photon transition in
helium-like uranium \cite{4}. In this contribution we extend our
studies to systems with up to five electrons and we adopt the
ingenious ideas proposed by Botz, Bru{\ss}, and Nachtmann \cite{20}
especially suited for the investigation of parity-violating effects in
storage rings.

The starting point for all such experiments is that, due to the
parity-violating exchange of neutral $Z$ bosons between nucleus and
electrons, every electron state is mixed with states of opposite parity. In
first order perturbation theory the coefficient $\eta$ of this admixture is
given by

\begin{equation}
\label{eins1}
\eta = \frac{\langle i | \frac{G_F}{2 \sqrt{2}} 
             ( 1- 4 \sin^2 \vartheta_{\rm W} - \frac{N}{Z} )
                              \rho \gamma_5   | f \rangle}{E_i - E_f}
\qquad,
\end{equation}
where $G_F$ denotes Fermi's constant, $\vartheta_W$ the Weinberg angle, $N$
the neutron number, $Z$ the proton number, and $\rho$ the nuclear density
normalized to $Z$. From this formula we see why heavy ions with few electrons
left in inner shells are good candidates for investigating parity
non-conservation effects: The admixture coefficient $\eta$ is very large
(typically orders of magnitude larger than for usual, neutral atoms) due to
the big overlap between the nucleus and the electron states. The other factor
that can make  $\eta$ large is the energy difference between the two mixing
electronic states $i$ and $f$ that ought to be very small. Therefore we are
especially interested in level crossings of electron states with the same
spin but opposite parity.

It was pointed out in \cite{5} that equation (\ref{eins1}) has to be modified
by radiative corrections, the weak charge $Q_W$ included in (\ref{eins1})
changes according to

\begin{equation}
\label{zwei2}
Q_W = Z - 4Z \sin^2 \vartheta_{\rm W} - N \qquad \longrightarrow \qquad
     \rho_{PV}'( Z - 4Z\kappa_{PV}' \sin^2 \vartheta_{\rm W} - N)
\quad.
\end{equation}
Here $\rho_{PV}'$ and $\kappa_{PV}'$ are constants that arise from the
radiative corrections mentioned above. The crucial point is, that they depend
on the masses of the particles involved in the radiative processes,
especially the top quark and the Higgs boson. As it seems now that there is
evidence for the top quark to exist, it should be from a theoretical point of
view possible to determine from $\rho_{PV}'$ and $\kappa_{PV}'$ the value of
the mass of the Higgs boson that makes the standard model renormalizable,
thus giving important guidance to identify this particle in high energy
experiments.

In section 2, we will discuss uranium with two to five electrons as a model
for other heavy ions reaching from gold to plutonium. In section 3 we will
discuss the possibility of level crossing in compound heavy ions and finally
in section 4 we will investigate the possibility of polarization rotations in
heavy ions.

\section{Heavy Ions with two to five Electrons in inner Shells}

As to an experiment with heavy ions with few inner shell electrons we
have to give a criterion under which we can judge feasibility of such
an experiment. As such a criterion we should compare the $\eta$ values
of the systems regarded here with the $\eta$ value of the helium-like
uranium system discussed in \cite{4}, i.e.,  $\eta \approx 10^{-6}$,
when taking the energy difference to $\Delta E = 1$ eV.  Even this
relatively high value of $\eta$ left the proposed experiment beyond
the scope of experimental feasibility for the set up discussed there.

We get a second restriction by the following consideration. 
If, for example, the interesting electron states are excited during
the stripping process of the ion in a stripping foil, 
then for any realistic experiment the experimental set up should be
placed a little distance behind this foil, lets say one meter. Then
the lifetime of these excited states should be long enough to survive
this one meter of flight.  Taking into account a time dilation factor
of about 5 for an ion accelerated to 5 GeV per nucleon, the lifetime
should be larger than $\approx 10^{-9}$ s. This would be an optimal value,
but also a lifetime of $10^{-10}$ s would probably do, corresponding to a
distance of 10 cm.

We furthermore consider only the lowest lying electron states that
offer a possibility for a parity-violation experiment.  As the parity
admixture is proportional to the overlap of the electron states in
question with the nucleus, this admixture should become very large for
low lying states, if the energies are sufficiently degenerate.

We should state that these criterions do not rule out all imaginable
experiments. It could be possible for example to store ions in an ion
trap and to generate the excited state by a laser beam, maybe by a
laser that still has to be invented or that will be available in a few
years, then the question of the lifetime of the electron states may be
superfluous. Also there could be other electronic configurations in
the ions studied here or in ions having a few more electrons, with
suitable properties. The relativistic corrections can lead to very
rich structures, with level crossings and metastable states, which have
just begun to be explored (see for example \cite{223,ind96}) and can lead to
 increased sensitivity to e.g., electric quadrupole hyperfine interaction \cite{725}.

The systems we are interested in are highly charged heavy ions for
which two states with equal angular momentum $J$, but opposite parity
have similar energy. We have investigated the binding energies of the
lower lying levels of ions with two to five electrons, to identify the
most promising candidates. When not available from the literature,
energies were calculated with the Multiconfiguration Dirac-Fock
Program (MCDF) published by Grant et {\it al.} \cite{6}, which provide
relativistic correction as well as one-electron QED corrections and
approximate, although inaccurate, many-body QED corrections. For
helium-like systems however, we can use very precise MCDF or
relativistic Configuration-Interaction (RCI) calculations including
correlation and QED effects.

For each electron configuration we show as an example the results for
uranium. There are no noticeable qualitative differences for other
heavy ions down to gold as is graphically shown for the interesting
electron states, except for the two-electron $1s2s~^1S_0\rightarrow
1s2p~^3P_0 $ case for which two crossings at $Z\approx 62$ and
$Z\approx 92$ occur.  As in this section we do only exploratory work,
we do not claim a precision much better than a few eV, except for
two-electron systems. Lifetimes are calculated in the single LS
configuration approximation. We take the inverse of the main transition
probability to be the lifetime of the respective state, neglecting hereby
other contributions of lower order. The parity admixture coefficient in this
second section is determined only for the main electron state and therefore
also gives only the order of magnitude.

\subsection{Two-electron ions}

Extensive calculations of two-electron ions binding energies have
appeared in the literature over the past 10 years
\cite{10,igd87,dra88}. In figure \ref{fig:he_cross}, we plot the
$1s2s~^1S_0\rightarrow 1s2p~^3P_0 $ energy difference as calculated in
\cite{10} and \cite{iab95}. The first one is an all-order Relativistic
Many-Body Perturbation theory (RMBPT) calculation, which uses Ref.~
 \cite{dra88} two-body QED corrections. The second calculation is a
MCDF calculation done along the line of \cite{igd87,ind95}, which uses
the Welton model for two-body self-energy corrections, experimental
nuclear size when available and includes finite-nuclear size correction
to the self-energy \cite{mas93}. The energy separation between
$1s2s~^1S_0$ and $1s2p~^3P_0 $ is plotted in Fig.\ref{fig:he_cross} as
a function of Z. In order to show how this level crossing happens we
show in detail the contributions to the energy separation at $Z=62$ and 92
in table \ref{tab:hecross}.  It should be noted that this crossing 
mostly involves the
interplay between magnetic energy and QED correction contributions.

With this new energy determination the parity admixture $|\eta| \approx
5\times 10^{-6} \mbox{eV}/(\Delta E)$ \cite{4} would be enhanced by a
factor of 3. For the experimental set up discussed in \cite{4} with
the detection of a laser-induced two-photon transition, the laser
intensity required would still be unrealistically large, of order
$10^{21}$ W/cm$^2$ (presently only lasers up to an intensity of
$10^{17}$ W/cm$^2 - 10^{18}$ W/cm$^2$ exist.). The main problem in
this context is that the heavy ions are only available in the form of
a rapid ion beam and that the only possibility to excite the
interesting electron states is by means of the stripping process.

One hope to improve the situation is to study different isotopes to
see if one could still reduce the energy difference. Fig.\
\ref{fig:he_iso} shows that by chosing suitable isotopes the
degeneracy can be improved. Only the Coulomb energy is modified due to
the change in nuclear radius. For uranium the energy separation does
cancel between isotope 233 and 234, within the present
calculation. One should keep in mind, however, that the present
calculation as well as the one in \cite{10} are not precise enough for
finding exactly at which atomic number and for which isotope the
crossing occurs. The main uncertainty is in the self-energy
screening. In table \ref{tab:hecross} the self-energy screening is
evaluated with the Welton model\cite{igd87}, which has been proven to
be rather accurate \cite{1065}, but which is not ab-initio. In
\cite{10} Drake's screening calculations, which are more adapted to
low $Z$ are used. If one uses ab-initio QED calculations \cite{iam91},
one gets a larger screening. However \cite{iam91} did not include
relaxation, which seems to be sizable for the $1s2s~^1S_0$ state. For
uranium the Welton model with relaxation gives 4.29 eV, while the
result from \cite{iam91} is only 1.08 eV. It has been shown on other
systems that the Welton model should not be wrong by more than 10\%
for this atomic number, while it can be good to 1\% at lower Z
\cite{1065}. One should note also that higher order radiative
corrections (of order $\alpha^2$, i.e., of order $\alpha$ with respect
to the one electron self-energy), and QED corrections to the
two-photon exchange diagrams \cite{bmj93} have not been
evaluated. Both corrections could be as large as 0.5 eV. The position
of the crossing point as well as the smallest energy which can be
obtained is thus very uncertain. Also it should be remembered that if
the energy separation is too small it may be difficult to find a laser
to excite the two-photon transition.

\subsection{three to five-electron ions}

The characteristic feature of the lithium-like uranium (cf. table
\ref{tab:lilike}) is the fact, that already the ground state and the
first excited state fulfill the main conditions of a parity-violation
experiment, i.e., they have the same angular momentum and opposite
parity. Moreover the lifetime of the first excited state lies in the
range of $10^{-10}$ seconds.  Very sophisticated calculation of the
ionization energies in lithium-like uranium including discussion on
nuclear effects can be found in \cite{210,11,920}. Complete
calculations with relativistic correlation energy and radiative
corrections for lower atomic numbers can be found in
Refs\. ~\cite{423,326,74} Unluckily, between these two energy states
there is a wide energy gap that reduces the magnitude of the parity
admixture, which is in rough approximation about $\eta = 1.4 \times
10^{-8}$. We shall discuss a scheme of detecting parity-violation in
lithium-like atoms in section 3.

Fig.\ \ref{fig:lith} shows that the $Z$-dependence of the energy
difference of the first two electron states is nearly linear for
atomic numbers ranging from $79\leq Z \leq 92$, such that no element
can be found for which the situation would be substantially different.

The beryllium-like ions case is comparable to the lithium-like
case. The first two electron levels are in principle suitable for
parity admixture experiments. The lifetime of the first excited state
is very large and depends crucially on the spin of the nucleus
\cite{mpi93}. In the case of an even-even nucleus, e.g., uranium 238
the lifetime is dominated by a two-photon $E1M1$ transition which is
in general very slow ($10^7s$ for $Z=82$\cite{mpi93}),
and can therefore be treated to be infinity in comparison with the 
lifetime of the next higher levels. 
In the case of uranium 235 the nucleus has an angular
momentum of 7/2, and due to hyperfine mixing of electron orbitals the
lifetime is severely reduced to $8.562 \times 10^{-5}$ s \cite{mpi93}.

As a model for beryllium-like heavy ions we tabulate the energy and
lifetime of the lower level of beryllium-like uranium in table
\ref{tab:belike}. In order to get a reasonable precision both the
ground state and the $1s^22p_{1/2}^2$ are calculated as the lower and
intermediate level of the $1s^22s^2+1s^22p_{1/2}^2+1s^22p_{3/2}^2 J=0$
configuration set, because intrashell correlation is very large in
that case. As in the lithium-like case the energy gap between the
mixing levels is large, leading to a parity admixture of about $|\eta|
\approx 2.4\times 10^{-8}$.

For a five-electron system we examine again uranium ions (cf. table
\ref{tab:bolike}). The first two
electronic levels are in principle usable for a parity-violation
experiment, but the comparatively short lifetime of the first excited
state and the small admixture of only $|\eta| \approx 9.4 \times
10^{-9}$  make this system completely unattractive.  We shall discuss
in the following therefore mainly lithium-like ions.

 No level crossing was found for  $78 \leq Z
\leq 96$ in any of the three, four and five-electron systems.

\section{Lithium-like heavy ions with high Z and N}
In this section we study super-heavy lithium-like ions.  It is
interesting to check, how the situation would change if $Z$ would be
increased beyond the existing periodic system. Such high-$Z$ systems
can be formed for a short time in heavy-ion collisions.  Here we treat
the high-$Z$ system as an ordinary atom with the charge $Z=Z_1+Z_2$
being just the sum of its components. While the energy difference
$E(1s^2 2p ~^2P_{1/2}) - E(1s^2 2s ~^2S_{1/2})$ is nearly linearly
increasing in the range from $Z=79$ to $Z=94$ it again decreases in
the higher $Z$ region and has a crossing point at $Z_{united} \approx
122$.  
This effect is due to the relativistic contraction of the $2p_{1/2}$
wave function  which dominates over all other contributions
for very large $Z$. For further increasing $Z$ the $2p_{1/2}$ state, 
being below the $2s_{1/2}$ state, reaches
the the negative energy continuum \cite{GMR85}. 
We used Desclaux's code to evaluate a number of systems for
$104 \leq Z \leq 128$, with self-consistent magnetic interaction
\cite{ind86}, vacuum polarization of order $\alpha (Z \alpha)$,
$\alpha (Z \alpha)^3$ and $\alpha^2 (Z \alpha)$, self-energy
extrapolated from Mohr's values and corrected for finite nuclear
size. For this to be valid, however we had to limit ourselves to $Z <
137$. It happens that the interesting region lies well inside this
boundary. From table \ref{tab:supheav}, one can see how for such high
$Z$ values the two interesting lithium-like states cross around the
united charge number $Z_{united} \approx 122$.  We analized only
symmetric collision systems, which are parity even, provided their
charge states are equal.

\section{Polarization rotations}
This section follows the analysis given in \cite{20} by G.~W.~Botz,
D.~Bru{\ss} and O.~Nachtmann. We follow here their notations.  The
energies, lifetimes, Stark and parity admixture coefficients were
calculated with the Multi-Configuration Dirac Fock package from
\cite{15}. To make this paper self contained let us shortly repeat
some of the basic arguments of \cite{20}.

The atomic system we are interested in is a lithium-like ion that has
a nonzero nuclear angular momentum. For simplicity we take the nuclear
angular momentum $I= 1/2$ and look at the first four electron states
(cf. Fig.~\ref{fig:hyp}). The situation for ions with other nuclear
angular momentum is completely the same except that other numbers for
the total angular momentum $F$ have to be inserted (The formalism
could also be applied to the boron-like case where we look at 
boron-like uranium 235 that has $I=7/2$.).  The experimental situation in
which we like to place this system is shown in Fig. \ref{fig:exp}.

The lithium-like ion moves in the 1-direction of our coordinate
system. This ion is moving through alternating electric fields of
width $x_1$, at a distance of $x_2$. The electric fields point in the
positive and negative 3-direction.  The moving ion sees a magnetic
field due to the boost, but as this field is even under parity
transformation we can neglect it.
\newline
The arrangement has still one symmetry operation $\hat R$ under which
it is invariant and this is a combination of parity transformation and
rotation about $\pi$ around the 2-axis.  Together, this gives a
reflection with respect to the 1-3 plane.
\begin{equation}
R : \left( \begin{array}{c}
             x_1 \\
             x_2 \\
             x_3 \\
         \end{array}    \right)
\longrightarrow
\left( \begin{array}{c}       
             x_1 \\
            -x_2 \\
             x_3 \\
         \end{array}     \right)
\end{equation}
\begin{equation}
\hat R = e^{i \pi \hat F_2} \cdot \hat P \quad.
\end{equation}
It is clear that the angular momentum states $|F,F_3\rangle$ are in general no 
eigenstates of this operation. But from 
\begin{eqnarray}
e^{i \pi \hat F_2}|F,F_3\rangle &=& \sum_{F_3'} |F,F_3'\rangle \langle F,F_3'|e^{i\pi \hat F_2}|F,F_3 \rangle
\nonumber \\
 &=&  \sum_{{F_3}'} |F,{F_3}'\rangle D_{{F_3}',F_3}^{(F)*}(0,-\pi,0)
\nonumber \\
 &=&   \sum_{{F_3}'} |F,{F_3}'\rangle d_{F_3,{F_3}'}^{(F)}(\pi)
\nonumber \\
 &=&   \sum_{{F_3}'} |F,{F_3}'\rangle (-1)^{F-{F_3}'} \delta_{{F_3}',-F_3}
\nonumber \\
 &=&    (-1)^{F+F_3} |F,-F_3\rangle  
\end{eqnarray}
it is easily seen that states with $F_3=0$ still are eigenstates
of the reflection symmetry operator, and for simplicity we will constraint our
considerations to those states.
\newline
\newline
This reflection symmetry is destroyed by the weak interaction of the 
electron with the nucleus which adds to the atomic Hamiltonian the
terms
\begin{eqnarray}
H_{PV} &=& H_{PV}^{(1)} + H_{PV}^{(2)} \quad,
\nonumber \\
\nonumber \\
H_{PV}^{(1)} &=& - \frac{G_F}{\sqrt 2} \int d^3 x 2 g_A^e 
                \overline e({\bf x}) \gamma^\lambda \gamma_5 e({\bf x})
                \left( \sum_q g_V^q \overline q ({\bf  x}) \gamma_\lambda q({\bf x}) \right),
\nonumber \\
\nonumber \\
H_{PV}^{(2)} &=& - \frac{G_F}{\sqrt 2} \int d^3 x 2 g_V^e
                \overline e({\bf x}) \gamma^\lambda e({\bf x})
                \left( \sum_q g_A^q \overline q ({\bf x}) \gamma_\lambda \gamma_5 q({\bf x}) \right).
\end{eqnarray}
Here $q$ runs over all quarks, $G_F$ is Fermi's constant and 
$g_{A,V}^{e,q}$ denotes the neutral current coupling constants for the
quark flavour $q$ or the electron $e$, respectively. Both terms together have
no defined parity and consequently no defined quantum number  
according to the reflection symmetry operation $\hat R$.\newline
\newline   
\newline
On its flight the ion stays for the time
$t_1$ in the Stark field and during the time $t_2-t_1$
outside of it. 
\newline 
Following 
essentially the notation of \cite{20} we get for the transition
amplitude during the time $t_1$, 
in the case that there is no change in angular momentum
\begin{equation}
f_{F,F_3;F,F_3}(t_1) = \exp \left\{ -i E(2 \hat S,F)t_1 
                      -i \tilde \kappa_{F,F_3} 
                    \left( \frac{\sqrt 3 {\cal F}}{L} \right)^2 L t_1 -  
                           \kappa_{F,F_3}
                    \frac{1}{2}
                    \left( \frac{\sqrt 3 {\cal F}}{L} \right)^2  \Gamma t_1 
                    \right\}
\end{equation}           
In this formula we take $E(2 \hat S,F)$ to be the energy of the $2S$ hyperfine states, 
perturbed by the parity-violating weak interaction denoted by the hat over
the $S$. ${\cal F}$ is the electric stark field ${\cal E}$ 
multiplied with $e$ and the Bohr radius:
\begin{equation}
{\cal F} = \frac{e}{Z \alpha m_e} {\cal E} \qquad.   
\end{equation}
$L = E_{2S_{1/2}}-E_{2P_{1/2}}$ is the energy difference of the two 
electron states of opposite parity considered in Fig.~\ref{fig:hyp} 
and $\Gamma$ the decay constant of the $2P_{1/2}$ state mentioned above.
Here the hyperfine splitting is neglected because of its relative smallness.
The $\kappa$'s are perturbative constants that give
the admixtures due to the quadratic Stark effect.
\newline
\newline
In the case that there is a transition between the angular momentum states the
amplitude is proportional to the applied electric field i.e., e.g.
\begin{equation}
f_{1,0;0,0} \sim {\cal F}  \quad.
\end{equation}
The total transition amplitude for an ion flying through one capacitor and
the subsequent free drift length is given by
\begin{equation}
g_{F',F'_3;F,F_3} = e^{-iE(2 \hat S,F')(t_2 - t_1)} f_{F',F'_3;F,F_3}(t_1)
\quad.
\end{equation}
For an experimental set up with $K$ capacitors 
the amplitude for the $R$ symmetry 
violating transition $|F= 0 F_3 = 0 \rangle \rightarrow | F = 1 F_3 =0 \rangle$
is: 
\begin{eqnarray}
f^{(K)}_{1,0;0,0} &=& g_{1,0;0,0} \sum_{k = 0}^{K-1} 
                       g_{0,0;0,0}^k g_{1,0;1,0}^{K-k-1}
\nonumber \\
\nonumber \\
     &=& g_{1,0;0,0} \; g^{K-1}_{1,0;1,0} \frac{1 - \left( \frac{g_{0,0;0,0}}{g_{1,0;1,0}} \right)^K}
                        {1 - \left( \frac{g_{0,0;0,0}}{g_{1,0;1,0}} \right)}
\quad.
\end{eqnarray}
The basic idea is now to make the absolute value of such a transition amplitude large.
To this end, with the definitions given before one can express first
\begin{eqnarray}
\label{4.12}
\frac{g_{0,0;0,0}}{g_{1,0;1,0}} &=& \exp \left\{ +i \left[  A t_2 
                                        - (\tilde \kappa_{0,0} - \tilde \kappa_{1,0}) 
                                         \left( \frac{\sqrt 3 {\cal F}}{L} \right)^2L t_1 \right] \right.
\nonumber \\
\nonumber \\
   & & \left. \quad - \frac{1}{2} ( \kappa_{0,0} - \kappa_{1,0}) 
 \left( \frac{\sqrt 3 {\cal F }}{L} \right)^2 \Gamma t_1 \right \} \quad.
\end{eqnarray}
Here $A = E(2\hat S,1)- E(2 \hat S,0)$ denotes the energy difference due to hyperfine splitting of the
2S electron orbitals. This very expression can be made real by a suitable choice of the length of the free drift
space so that the condition
\begin{equation}
\label{3}
A t_2   - (\tilde \kappa_{0,0} - \tilde \kappa_{1,0})
\left( \frac{\sqrt 3 {\cal F}}{L} \right)^2L t_1 = 2 \pi n 
\end{equation}
holds.
We will come back to this later. With the above choice of $t_2$ we can get for the absolute value of the 
amplitude $f^{(K)}_{1,0;0,0}$ the expression
\begin{eqnarray}
|f^{(K)}_{1,0;0,0}| &\sim& (\sqrt 3 {\cal F} t_1) 
|g_{1,0;1,0}|^{K} \frac{1 - \left( \frac{g_{0,0;0,0}}{g_{1,0;1,0}} \right)^K}
                         {1 - \left( \frac{g_{0,0;0,0}}{g_{1,0;1,0}} \right)}
\nonumber \\
\nonumber \\
		    &= &   \frac{1}{2 \sqrt{Q}} \quad.
\end{eqnarray}
Here we have assumed $K \gg 1$.
Now the aim is to maximize $|f^{(K)}_{1,0;0,0}|$ which is the same as to minimize $Q$. This quantity $Q$ plays an 
important role in this connection because as is shown in \cite{20} $Q$ is a measure for the polarization rotation 
of the ion flying through the capacitor arrangement as at $t=0$ there is no component of angular momentum $F$
parallel to the direction of flight.
\begin{equation}
|{\bf e_1} \cdot {\bf \hat F} (K t_2)| \sim \frac{1}{2\sqrt{Q}} \quad. 
\end{equation}
For definiteness we discuss the case of a pair of states with $F=0$ and $F=1$.  We abbreviate
\begin{eqnarray}
\label{2}
x &=& \frac{1}{2} ( \kappa_{0,0} - \kappa_{1,0}) 
       \left( \frac{\sqrt{3} {\cal F}}{L} \right)^2 \Gamma t_1 K \quad,
\nonumber \\
\nonumber \\
\kappa &=& \frac{2 \kappa_{1,0}}{\kappa_{0,0} - \kappa_{1,0}} \quad,
\nonumber \\
\end{eqnarray}
and use as independent variables $x$ and $K$. We get up to factors independent of $K$ and $x$ 
\begin{equation}
Q \sim \frac{K}{x}e^{\kappa x } \cdot \frac{( 1- e^{-x/K})^2}{(1-e^{-x})^2}
\quad.
\end{equation}
Let us assume $K$ to be large, then $Q$ is antiproportional to the number of capacitors $K$.
We now treat $K$ as a fixed number and then look for the minimum 
of $Q$ as a function of $x$. As $K \gg 1$ the formal minimum of $Q$ is 
obtained for $x\ll 1$ such that in the vicinity of the minimum one has 
\begin{equation}
Q \sim \frac{e^{\kappa x}}{xK} \quad \rightarrow \qquad x_{\rm min} \sim \frac{1}{\kappa}
\quad.
\end{equation}
At the minimum the quantity ${\cal F}$, essentially the electric
field ${\cal E}$,  is determined by 
\begin{equation}
\label{6}
\left( \frac{\sqrt{3} {\cal F}}{L} \right)^2 =
\frac{1}{\kappa_{1,0} K \Gamma t_1}
\quad.
\end{equation}
We shall discuss below that this optimal situation cannot be reached for 
the ions considered here.
The derivation of these equations has been done for a pair of
atomic states $F=0,F=1$. But there is no principal difference
for other combinations like $F=3,F=4$, which is considered here for
boron-like uranium. %
\newline
\newline
While the formulae are just the same as derived in \cite{20}
 the quantities involved are quantitatively very different. 
Various large factors appear both in favor and in disfavor of the
heavy-ion system and there is no simple way to estimate the
relative size of the effect. 
We shall present the 
numerical results for $U^{235}$ in table \ref{table:AGL}. It will turn out 
that also some light ions might be of interest. 
Therefore we also add 
to table \ref{table:AGL} the results for  the three lithium-like systems 
$Be^+$, $B^{2+}$ and $C^{3+}$. Their atomic properties are shown
in tables \ref{Be9}, \ref{Bo11} and \ref{Ca13}. The atomic properties
of $~^{235}U$ are shown in tables 
\ref{tab:235U}, \ref{MEUr235} and \ref{HFUr235}.
For the calculation of the $\kappa$ coefficients we use perturbation
theory:
\begin{eqnarray}
\tilde \kappa_{F,F_3} \left( \frac{\sqrt{3} {\cal F}}{L} \right)^2 L 
    & = & \sum_{n \neq m} 
          \frac{|\langle n| e E z | m \rangle|^2}{E_m - E_n}\quad,
\nonumber \\
\nonumber \\
 \kappa_{F,F_3} \left( \frac{\sqrt{3} {\cal F}}{L} \right)^2 \Gamma
    & = & \sum_{n \neq m}
           \left|\frac{\langle n| e E z | m \rangle}{E_m - E_n}\right|^2 \Gamma_n
\quad.
\end{eqnarray}
Here $m$ denotes the state with the quantum numbers $F,F_3$ and $n$ the
other admixing states. Solving this for the $\kappa's$ and using the 
Wigner Eckart, 6j- and 9j- theorems one gets
\begin{eqnarray}
\tilde \kappa_{F,F_3} &=& \frac{1}{3} {(2F+1)}
                          \sum_n 
                          {(2F_n+1)} \frac{L}{E_m-E_n}
\nonumber \\
\nonumber \\
&&                  \frac{|\langle n j_n||z||m j_m \rangle|^2}{r_B(Z)^2}
                       \left(
                          \begin{array}{ccc}
                            F_n & 1 & F \\
                           -F_3 & 0 & F_3 \\
                          \end{array} \right)^2
                          \left\{
                          \begin{array}{ccc}
                            F_n & j_n &  I  \\  
                            j   & F   &  1  \\
                          \end{array} \right\}^2 \quad,
\nonumber \\
\nonumber \\
\kappa_{F,F_3} &=&      \frac{1}{3} {(2F+1)}
                          \sum_n
                         {(2F_n+1)} 
                          \left(\frac{L}{E_m-E_n}\right)^2
\nonumber \\
\nonumber \\
&&              \frac{|\langle n j_n||z||m j_m \rangle|^2}{r_B(Z)^2}
                          \left(
                          \begin{array}{ccc}
                            F_n & 1 & F \\  
                           -F_3 & 0 & F_3 \\
                          \end{array} \right)^2
                          \left\{
                          \begin{array}{ccc}
                            F_n & j_n & I \\
                            j   & F   & 1 \\  
                          \end{array} \right\}^2  \quad.         
\end{eqnarray}
Here $r_B(Z) = 1/(Z \alpha m_e)$. 
The point is now that the kappa coefficients only deviate by the
small energy differences that are due to the hyperfine splitting. 
In table \ref{table:kappa} we
show the $\kappa$ values for $Be^+$, $B^{2+}$, $C^{3+}$ and
$U^{87+}$.
Together with the numerical values of the hyperfine splitting and the Stark 
matrix elements, which are given in tables \ref{Be9}--\ref{HFUr235}, 
we can calculate the interesting 
expressions for the polarization rotation effects.
Let us first start in the same way as in \cite{20} and 
analize the situation for the minimal $Q$. 
It will turn out, that this assumption would imply unrealistically
large electrical fields resulting from:
\begin{equation}
t_2 = \frac{\tilde \kappa_{0,0} - \tilde \kappa_{1,0}}{\kappa_{1,0}}
      \frac{L}{\Gamma  A} \left[ \frac{\hbar}{e}\right] \frac{1}{K}
\qquad (n = 0)\quad. 
\end{equation}
From (\ref{4.12}) we get the requirement for the individual effects to
add.
\begin{equation}
\left[A t_2 - \frac{\tilde \kappa_{0,0} - \tilde \kappa_{1,0}}{\kappa_{1,0}}
\frac{L}{K\Gamma} \right] = 2 \pi n \quad. 
\end{equation}
This implies that the deviation $\delta t_2$ in $t_2$ should be smaller
than
\begin{equation}
\delta t_2 < \frac{1}{100} \frac{1}{A} \left[ \frac{\hbar}{e} \right] \quad.
\end{equation}
For the time $t_1$, which gives the length of the capacitor we are required
to take $t_1 \leq t_2$, but there are no other constraints.
To make the required electric field small one has to chose $t_1$ large
[see Eq. (\ref{4.24}) below], so
we take $t_1 = t_2/2$.
Finally from the relation 
\begin{equation}
\label{4.24}
\left( \frac{\sqrt{3} {\cal F}}{L} \right)^2 = \frac{1}{\kappa_{1,0} \Gamma t_1}
        \left[\frac{\hbar}{e} \right] \frac{1}{K}
\end{equation}
one can well calculate the electric field. The terms in $[\dots]$ give
always the necessary factors for the translation into SI units.
The resulting numbers are given in table \ref{tab:txQ}. 
Here we always set $K=1$.
The values for other $K$ can easily be determined from the formulas above. 
Note that $K$ has to be chosen very large and that 
the electric field ${\cal E}$ is for the
choice $t_1 = t_2/2$ or for any choice $ t_1 \sim t_2$ independent of $K$.
Table \ref{tab:txQ} shows the results for $Be^+$, $B^{2+}$, $C^{3+}$ and    
$U^{87+}$.
The values for the electric field ${\cal E}$ are so unrealistically large
that such an experiment cannot be realized. The reason for the
large values of ${\cal E}$ is the fact that in atoms with more than
one electron the energy difference between the $2p_{1/2}$ and the $2s_{1/2}$
states 
is orders of magnitude larger than for hydrogen-like atoms
because the $2s_{1/2}-2p_{1/2}$ degeneracy is removed by the electron-electron
interaction.
\newline
\newline
We now proceed in the opposite direction. We take a realistic field 
${\cal E}$ and
other realistic values
\begin{eqnarray}
K &=& 1000 \quad, 
\nonumber \\
{\cal E} &= & 1000 \frac{V}{m} \quad,
\nonumber \\
t_1 & = & 1.0 \times 10^{-8} s \quad.
\end{eqnarray}
We then calculate 
\begin{equation}
x = \frac{1}{2}(\kappa_{0,0} - \kappa_{1,0})
    \left( \frac{ \sqrt{3} {\cal F}}{L} \right)^2 \Gamma t_1 K
\left[ \frac{e}{\hbar} \right]
\quad.
\end{equation}
As $x$ is very small we approximate
\begin{eqnarray}
Q &=& \frac{(\kappa_{0,0}- \kappa_{1,0}) \Gamma K}{8 L^2 t_1}
      \frac{1}{x} \left(1- e^{-\frac{x}{K}}\right)^2 (1-e^{-x})^{-2}
      \exp \left( \frac{2 \kappa_{1,0}}{\kappa_{0,0}  - \kappa_{1,0}}x \right)
\nonumber \\
\nonumber \\
  & \longrightarrow&
     \frac{\kappa_{0,0}- \kappa_{1,0}  }{8 L^2 t_1}
     \Gamma \frac{1}{xK} \left[ \frac{\hbar}{e} \right] \quad.
\end{eqnarray}
In this way we get the values of table \ref{tab:realistic}.
These values must be compared to that obtained in \cite{20} 
for hydrogen $Q_{\rm min} = 6.6 \times 10^{-9}$.
\section{Conclusions}
In principle it is obvious that heavy ions with few inner shell electrons 
offer a possibility to test the effects of parity admixture. This admixture
is in heavy ions orders of magnitude larger than in neutral atoms.


The ideal case is a parity-violation effect to a sizable extent
without applying any of the elaborated methods used in the cesium experiment
\cite{3}.
Then, the only chance to
get measurable parity admixtures is to find a pair of energy states
near the ground state with equal angular momentum but opposite parity
that is nearly degenerated with respect to its  energy.
Unfortunately there is no such pair of orbitals
in uranium with two to five electrons except for the
already known degeneracy in helium-like uranium. As the electron levels
do only change very slowly with $Z$ the same
is true for the neighboring heavy ions.\newline
The next step will consequently be a very detailed analysis
of the degeneracy in helium-like heavy ions including nuclear and isotopic
effects because here a level crossing must exist.
Level crossing also exists for compound nuclear reactions but here the
lifetime of the compound nucleus is too short to allow for 
atomic physics experiments. 
Looking for parity-violating spin rotations opened another perspective.
We showed, however, that the net effect
(value of $1/\sqrt{Q}$) for heavy ions is  
about thirty times weaker than for hydrogen. 
%
%


\begin{table}
\caption{Contributions to the $1s2s ^1S_0 \rightarrow 1s2p~^3P_0$
separation near the two crossing points. All units are given in eV.}
\begin{center}
\label{tab:hecross}
\begin{tabular} {lrrrrrrrr}
	      &$Z=62 $    &           &           &$Z=92$     &           &           \\
	      &$1s2p~^3P_0$&$1s2s~^1S_0$& Diff &$1s2p~^3P_0$&$1s2s~^1S_0$&
 Diff      \\
\hline  Coulomb       & -68868.56& -68861.61 &     -6.948& -165518.05& -165487.55&
  -30.50\\
Magnetic      &     38.30&     17.12&     21.17&     151.30&      66.36&      84.91\\
Retardation   &     -3.26&      1.30&     -4.56&     -10.09&       5.56&     -15.65\\
Mass Pol      &     -0.029&      0.00&     -0.03&     -0.04&       0.00&      -0.04\\
Correlation   &     -0.39&     -0.59&      0.20&     -1.02&     -1.18&       0.16\\
1e- Self-energ&     82.66&     95.16&    -12.50&     364.88&     420.68&     -55.80\\
2e- Self-energ&     -0.18&     -1.24&      1.06&     -1.15&      -5.44&       4.29\\
Uehling       &    -13.52&    -15.18&      1.67&     -96.13&    -108.71&      12.59\\
      &     -0.02&     -0.10&      0.087&      -0.28&      -0.89&       0.61\\
Wichman \& Kroll &      0.38&      0.42&     -0.04&       4.75&
5.28&      -0.53\\
Kallen \& Sabry       &     -0.11&     -0.12&      0.01&      -0.73&
-0.83&   0.09\\
Nuclear Pol.  &           &           &           &      -1.10&      -1.28&       0.18\\
Total energy  & -68764.71& -68764.83&      0.11& -165107.70& -165108.00&       0.30\\
\end{tabular}
\end{center}
\end{table}

\begin{table}
\caption{
Electron configuration of lithium-like uranium.
\label{tab:lilike}}
\begin{center}
\begin{tabular} {ccccc}
main conf.&parity& energy (eV)& lifetime (s)\\
\hline
 $1s^2 2s ~^2S_{1/2}$
      & $+$ & $-2.9424 \times 10^5$ & $\infty$\\   
 $1s^2 2p ~^2P_{1/2}$
      & $-$&  $-2.9395 \times 10^5$ &$1.0\times 10^{-10}$\\
 $1s^2 2p ~^2P_{3/2}$
      & $-$ & $-2.8978 \times 10^5$ &  $1.1\times 10^{-14}$ \\
 $1s^2 3s ~^2S_{1/2}$
      & $+$ & $-2.7545 \times 10^5$ & $4.9\times10^{-15}$\\
 $1s^2 3p ~^2P_{1/2}$
      & $-$ & $-2.7537 \times 10^5$ &  $4.6\times 10^{-16}$\\
\end{tabular}
\end{center}
\end{table}

\begin{table}
\caption{Electron configuration of beryllium-like uranium. \label{tab:belike}}
\begin{center}
\begin{tabular}{ccccc}
main conf. &parity& energy (eV)& lifetime (s)\\
\hline 
$1s^22s^2 ~^1S_0$ & $+$ &  -326604  & $\infty$ \\
$1s^22s2p ~^3P_0$ & $-$ &  -326345 & $ \infty$  for $U^{238}$\\
                      &&&$ \quad 8.56 \times 10^{-5}$ for $U^{235}$  \\
$1s^22s2p ~^3P_1$ & $-$ &  -326305 & $ 1.00\times 10^{-10}$ \\
$1s^22p^2 ~^3P_0$ & $+$ &  -325894  & $ 7.87\times 10^{-12}$\\
$1s^22s2p ~^3P_2$ & $-$ &  -322224 &$  3.37 \times 10^{-12}$\\
\end{tabular}
\end{center}
\end{table}

\begin{table}
\caption{Electron configuration of boron-like uranium.
\label{tab:bolike}}
\begin{center}
\begin{tabular}{ccccc}
main conf.& parity& energy (eV)& lifetime (s)\\
\hline 
    $1s^2 2s^2 2p ~^2P_{1/2}$
      &  $-$ &  $-3.5826 \times 10^5$ & $\infty$ \\
    $1s^2 2s 2p^2 ~^4P_{1/2}$ 
      & $+$ &  $-3.5785 \times 10^5$ & $5.2\times 10^{-11}$\\
    $1s^2 2s^2 2p ~^2P_{3/2}$
      &  $-$ &  $-3.5417 \times 10^5$ & $3.3 \times 10^{-12}$\\
    $1s^2 2s 2p^2 ~^4P_{3/2}$
      &  $+$&  $-3.5389 \times 10^5$ & $7.5\times 10^{-13}$\\
   $1s^2 2s 2p^2 ~^2D_{5/2}$
      &  $+$ &  $-3.5384 \times 10^5$ & $6.6\times 10^{-11}$\\
\end{tabular}
\end{center}
\end{table}

\begin{table}
\caption{Energies of the first two electron states in 
lithium-like heavy ions for high nuclear charges.
\label{tab:supheav}}
\begin{center}
\begin{tabular} {cccddd}
Name &  Z& A  &
    $E(1s^2 2s ~^2S_{1/2})$ &
    $E(1s^2 2p ~^2P_{1/2})$ &
    $\Delta$ (eV)  \\
&&& $(J^P=\frac{1}{2}^+)$ (eV) & $(J^P=\frac{1}{2}^-)$ (eV) &  \\
 \hline
Te+Te & 104 & 260  &   -396234.8  &     -395910.3  &     324.6\\
Ce+Ce & 116 & 280  &   -528168.1  &     -527979.6  &     188.5\\
Nd+Nd & 120 & 288  &   -581273.8  &     -581267.7  &       6.1\\
Sm+Sm & 124 & 304  &   -640357.1  &     -640692.8  &    -335.7\\
Gd+Gd & 128 & 316  &   -706756.4  &     -707698.3  &    -941.9\\
\end{tabular}
\end{center}
\end{table}

\begin{table}
\caption{Hyperfine splitting(A), level width($\Gamma$) 
and level separation (L) for selected ions. \label{table:AGL}}
\begin{center}
\begin{tabular}{ccccc}
Ion             & $Be^{+}$ &$B^{2+}$&$C^{3+}$&$U^{87+}$ \\
\hline
Isotope         & Be 9 & B 11 & C 13 & U 235 \\
lower state     &  2S1/2 &2S1/2 &2S1/2 & 1s2 2s2 2p J=1/2 \\
upper state     &  2P1/2 &2P1/2 & 2P1/2 &1s2 2s 2p2 J=1/2 \\
 I              &  3/2 &3/2 &  1/2 &  7/2 \\
A [eV]          & $1.71414 \times 10^{-6}$
                & $1.04438 \times 10^{-5}$ 
                & $8.44760 \times 10^{-6}$
                & $1.796   \times 10^{-2}$ \\
$\Gamma$ [eV]   & $7.79467 \times 10^{-8}$
                & $1.30629 \times 10^{-7}$
                & $1.81121 \times 10^{-7}$
                & $3.11949 \times 10^{-5}$ \\
L [eV]          & $3.98910 \times 10^{+0}$
                & $6.05385 \times 10^{+0}$
                & $8.07181 \times 10^{+0}$
                & $4.0302  \times 10^{+2}$ \\
\end{tabular}
\end{center}
\end{table}

\begin{table}
\caption{Atomic structure for lithium--like $~^9Be$.
\label{Be9}}
\begin{center}
\begin{tabular}{llr}
{\quad $2p_\frac{1}{2}-2s_\frac{1}{2}$ PNC matrix element}&&
--5.6810174$\times 10^{-15}$ eV \\
{\quad $2p_\frac{1}{2}-2s_\frac{1}{2}$ energy difference}&&
3.9891026$\times 10^{+00}$ eV \\
{\quad $2p_\frac{1}{2}-2s_\frac{1}{2}$ lifetime (length)}&&
8.4443912$\times 10^{-09}$  sec\\
{\quad $2p_\frac{1}{2}-2s_\frac{1}{2}$ lifetime (velocity)}&&
7.9463913$\times 10^{-09}$  sec\\
{\quad $2p_\frac{1}{2}-2s_\frac{1}{2}$ Stark--element}&&
.76491$ \times 10^{+00}$  a.u. \\
\hline
2p 1/2 F=2 &   total hyperfine matrix element: &   --1.1844518555$\times 10^{-07}$ eV\\
	  &    Bohr-Weisskopf correction:     &    5.9033102475$\times 10^{-15 }$ eV\\
	  &                 total:  &             --1.1844517964$\times 10^{-07 }$ eV\\
&&\\
2p 1/2 F=1 &  total hyperfine matrix element: &    1.9740864258$\times 10^{-07 }$ eV \\
	  &  Bohr-Weisskopf correction:      &   --9.8388504126$\times 10^{-15 }$ eV \\
	  &                 total:           &    1.9740863274$\times 10^{-07 }$ eV \\
&&\\
2s 1/2 F=2 &  total hyperfine matrix element:&    --6.4289041847$\times 10^{-07 }$ eV \\
	   &   Bohr-Weisskopf correction:    &     8.8568061986$\times 10^{-11 }$ eV \\
	   &                total:           &    --6.4280185040$\times 10^{-07 }$ eV \\
&&\\
2s 1/2 F=1 &  total hyperfine matrix element:&     1.0714840308$\times 10^{-06 }$ eV \\
	   &   Bohr-Weisskopf correction:    &    --1.4761343664$\times 10^{-10 }$ eV \\
	   &                total:           &     1.0713364173$\times 10^{-06 }$ eV \\
\end{tabular}
\end{center}
\end{table}
\begin{table}
\caption{Atomic structure for lithium--like $~^{11}B$. \label{Bo11}}
\begin{center}
\begin{tabular}{llr}
{\quad $2p_\frac{1}{2}-2s_\frac{1}{2}$ PNC matrix element}&&
--2.7316505$\times 10^{-14}$ eV \\
{\quad $2p_\frac{1}{2}-2s_\frac{1}{2}$ energy difference}&&
6.0538537$\times 10^{+00 }$ eV \\
{\quad $2p_\frac{1}{2}-2s_\frac{1}{2}$ lifetime (length)}&&
5.0387842$\times 10^{-09 }$ sec\\
{\quad $2p_\frac{1}{2}-2s_\frac{1}{2}$ lifetime (velocity)}&&
4.6813678$\times 10^{-09 }$ sec\\
{\quad $2p_\frac{1}{2}-2s_\frac{1}{2}$ Stark--element}&&
.52970$ \times 10^{+00 }$ a.u. \\
\hline
2p 1/2 F=2 &   total hyperfine matrix element: &   8.7104273520$\times 10^{-07 }$ eV\\
	  &    Bohr-Weisskopf correction:     &  --8.9772008878$\times 10^{-14 }$ eV\\
	  &                 total:  &             8.7104264543$\times 10^{-07 }$ eV\\
&&\\
2p 1/2 F=1 &  total hyperfine matrix element: &   --1.4517378920$\times 10^{-06 }$ eV \\
	  &  Bohr-Weisskopf correction:      &   1.4962001480$\times 10^{-13 }$ eV \\
	  &                 total:           &    --1.4517377424$\times 10^{-06 }$ eV \\
&&\\
2s 1/2 F=2 &  total hyperfine matrix element:&     3.9170586600$\times 10^{-06 }$ eV \\
	   &   Bohr-Weisskopf correction:    &      --6.4395425787$\times 10^{-10  }$ eV \\
	   &                total:           &     3.9164147057$\times 10^{-06  }$ eV \\
&&\\
2s 1/2 F=1 &  total hyperfine matrix element:&     --6.5284311000$\times 10^{-06 }$ eV \\
	   &   Bohr-Weisskopf correction:    &    1.0732570965$\times 10^{-09 }$ eV \\
	   &                total:           &     --6.5273578429$\times 10^{-06 }$ eV \\
\end{tabular}
\end{center}
\end{table}
\begin{table}
\caption{Atomic structure for lithium--like $~^{13}C$. \label{Ca13}}
\begin{center}
\begin{tabular}{llr}
{\quad $2p_\frac{1}{2}-2s_\frac{1}{2}$ PNC matrix element}&&
--8.7153650$\times 10^{-14 }$ eV \\
{\quad $2p_\frac{1}{2}-2s_\frac{1}{2}$ energy difference}&&
 8.0718138$\times 10^{+00 }$ eV \\
{\quad $2p_\frac{1}{2}-2s_\frac{1}{2}$ lifetime (length)}&&
3.6341015$\times 10^{-09 }$ sec\\
{\quad $2p_\frac{1}{2}-2s_\frac{1}{2}$ lifetime (velocity)}&&
3.3515826$\times 10^{-09 }$ sec\\
{\quad $2p_\frac{1}{2}-2s_\frac{1}{2}$ Stark--element}&&
.40514$ \times 10^{+00 }$ a.u. \\
\hline
2p 1/2 F=0 &   total hyperfine matrix element: &   --1.5529296692$\times 10^{-06  }$ eV\\
	  &    Bohr-Weisskopf correction:     &    3.0240473822$\times 10^{-13 }$ eV\\
	  &                 total:  &             --1.5529293668$\times 10^{-06 }$ eV\\
&&\\
2p 1/2 F=1 &  total hyperfine matrix element: &    5.1764322307$\times 10^{-07  }$ eV \\
	  &  Bohr-Weisskopf correction:      &   --1.0080157941$\times 10^{-13 }$ eV \\
	  &                 total:           &    5.1764312227$\times 10^{-07 }$ eV \\
&&\\
2s 1/2 F=0 &  total hyperfine matrix element:&    --6.3369865130$\times 10^{-06  }$ eV \\
	   &   Bohr-Weisskopf correction:    &     1.2851809538$\times 10^{-09  }$ eV \\
	   &                total:           &    --6.3357013320$\times 10^{-06  }$ eV \\
&&\\
2s 1/2 F=1 &  total hyperfine matrix element:&     2.1123288377$\times 10^{-06 }$ eV \\
	   &   Bohr-Weisskopf correction:    &    --4.2839365127$\times 10^{-10 }$ eV \\
	   &                total:           &     2.1119004440$\times 10^{-06 }$ eV \\
\end{tabular}
\end{center}
\end{table}
\begin{table}
\label{Ur235}
\caption{Atomic level structure for boron--like $~^{235}U$. \label{tab:235U}}
\begin{center}
\begin{tabular}{lcccc}
level &  binding energy ( eV) &    excitation energy (eV)\\
\hline
ground state &            --358233.01& ---- \\
1s2 2s 2p2 J=1/2 &       --357829.98&      403.02 \\
1s2 2s 2p2 J=3/2 &       --353861.72&      4371.29 \\
1s2 2s 2p2 J=5/2 &       --353818.09&      4414.92 \\
1s2 2s2 2p3/2   &        --354139.11&      4093.90 \\
1s2 2s 2p2 J=1/2  2nd &  --353712.14&      4520.87 \\
\end{tabular}
\end{center}
\end{table}

\begin{table}
\caption{ 1s2 2s2 2p J=$\frac{1}{2}$  --  1s2 2s 2p2 J=$\frac{1}{2}$ matrix 
elements in boron--like $~^{235}U$. \label{MEUr235}}
\begin{center}
\begin{tabular}{llcccc}
PNC matrix element       & 3.79$\times 10^{-06 }$ eV   \\
lifetime velocity gauge &  3.06$\times 10^{-11 }$ sec \\
lifetime length gauge   &  2.11$\times 10^{-11 }$ sec \\
Stark--element           & .29048$ \times 10^{-03 }$ a.u.\\
\end{tabular}
\end{center}
\end{table}
\begin{table}
\caption{Hyperfine structure in boron--like $~^{235}U$. \label{HFUr235}}
\begin{center}
\begin{tabular}{llr}
1s2 2s2 2p j=1/2  F=3  I=7/2
	  &   hyperfine:          &   1.02$\times 10^{-02 }$ eV\\
	  &   Bohr-Weisskopf:     &   --1.14$\times 10^{-04 }$ eV\\
	  &   total:              &   1.01$\times 10^{-02 }$ eV\\
&&\\
1s2 2s2 2p j=1/2 F=4 I=7/2
&  hyperfine:         &    --7.95$\times 10^{-03 }$ eV  \\
&  Bohr-Weisskopf:    &      8.83$\times 10^{-05 }$ eV \\
&  total:             &    --7.86$\times 10^{-03 }$ eV \\
&&\\
1s2 2s 2p2 j=1/2  F=3  I=7/2
&  hyperfine:          &    3.03$\times 10^{-02 }$ eV \\
&  Bohr-Weisskopf :    &  --9.83$\times 10^{-04 }$ eV \\
&  total:              &    2.94$\times 10^{-02 }$ eV \\
&&\\
1s2 2s 2p2 j=1/2  F=4  I=7/2
&  hyperfine:          &  --2.36$\times 10^{-02 }$ eV \\
&  Bohr-Weisskopf :    &    7.65$\times 10^{-04 }$ eV \\
&  total:              &  --2.28$\times 10^{-02 }$ eV \\
&&\\
\end{tabular}
\end{center}
\end{table}

\begin{table}
\caption{$\kappa$ values. \label{table:kappa}}
\begin{center}
\begin{tabular}{ccc}
Ion & $\kappa$-constants & numerical values\\
\hline
$Be^{+}$ &$\kappa_{1,0}, \tilde \kappa_{1,0}$ & $ 5.2007794253\times 10^{-1};\quad-5.2007778741\times 10^{-1}$\\ 
         &$\kappa_{2,0}, \tilde \kappa_{2,0}$ & $ 5.2007741321\times 10^{-1};\quad-5.2007752275\times 10^{-1}$\\  
&&\\
$B^{2+}$&$\kappa_{1,0}, \tilde \kappa_{1,0}$ & $3.8969640353\times 10^{-1} ;\quad -3.8969687978\times 10^{-1}$\\
        &$\kappa_{2,0}, \tilde \kappa_{2,0}$ & $3.8969804714\times 10^{-1} ;\quad -3.8969770158\times 10^{-1}$\\
&& \\
$C^{3+}$ &$\kappa_{0,0}, \tilde \kappa_{0,0}$ & $3.2827626620\times 10^{-1} ;\quad -3.2827654492\times 10^{-1}$\\
         &$\kappa_{1,0}, \tilde \kappa_{1,0}$ & $3.2827712174\times 10^{-1} ;\quad -3.2827697269\times 10^{-1}$\\
&& \\
$U^{87+}$ &$\kappa_{3,0}, \tilde \kappa_{3,0}$ & $3.9683205819\times 10^{-5}  ;\quad -3.9679966597\times 10^{-5}$\\
          &$\kappa_{4,0}, \tilde \kappa_{4,0}$ & $3.9669391256\times 10^{-5}  ;\quad -3.9673059448\times 10^{-5}$\\
\end{tabular}
\end{center}
\end{table}

\begin{table}
\caption{Characteristic values for selected ions. \label{tab:txQ}}
\begin{center}
\begin{tabular}{crrrr}
Ion & $Be^+$ & $B^{2+}$ & $C^{3+}$ & $U^{87+}$ \\
\hline
$t_2 [s]$       & $1.0000 \times 10^{-08} $ & $ 6.1594\times 10^{-09} $ & $ 4.5248\times 10^{-09} $ & $8.2441\times 10^{-11}$\\
$\delta t_2[s]$ & $3.8399 \times 10^{-12} $ & $ 6.3024\times 10^{-13} $ & $ 7.7917\times 10^{-13}$ & $3.6649\times 10^{-16}$\\
$t_1          $ & $5.0000 \times 10^{-09} $ & $ 3.0797\times 10^{-09} $ & $ 2.2624\times 10^{-09}$ & $4.1221\times 10^{-11}$\\
$x $            & $5.0888 \times 10^{-07} $ & $-2.1088\times 10^{-06} $ & $-1.3031\times 10^{-06}$ & $1.7412\times 10^{-04}$\\
$Q_{\rm min}  $ & $2.2789 \times 10^{-16} $ & $ 2.0174\times 10^{-16} $ & $ 1.8042\times 10^{-16}$ & $8.2689\times 10^{-20}$\\
${\cal E}[\frac{V}{m}]
              $ & $3.1371\times 10^{+11}  $ & $ 6.7668\times 10^{+11} $ & $1.1688\times 10^{+12} $ & $4.5952\times 10^{+16}$\\
\end{tabular}
\end{center}
\end{table}

\begin{table}
\caption{$x$ and $Q$ for realistic values of $K,\cal E$ and $t_1$.
\label{tab:realistic}}
\begin{center}
\begin{tabular}{rrrrr}
Ion & $Be^+$ & $B^{2+}$ & $C^{3+}$ & $U^{87+}$ \\
\hline
$x$ & $1.0341\times 10^{-20}  $ & $-1.4954\times 10^{-20} $ & $ -4.2159\times 10^{-21} $  & $ 2.0004\times 10^{-26}$\\
$Q$ & $2.0629\times 10^{-06} $ & $ 3.2232\times 10^{-06} $ & $  4.6414\times 10^{-06} $  & $ 1.0912\times 10^{-03}$\\
\end{tabular}
\end{center}
\end{table}


\clearpage
\figure
\caption{Energy difference of the two nearly degenerated electron states
as a function of atomic number. \label{fig:he_cross}}

\caption{Energy difference of the two nearly degenerated electron states
as a function of the mean-square nuclear radius, for $Z=92$. Value of the splitting
 for experimental nuclear size are represented by square dots.\label{fig:he_iso}}

\caption{Energy difference of the first excited state and the ground state
in lithium-like heavy ions from gold to plutonium $\Delta E = 
 E(1s^2 2p ~^2P_{1/2}) - E(1s^2 2s ~^2S_{1/2})$.\label{fig:lith}}

\caption{Hyperfine-splitting for the parity-mixed states.\label{fig:hyp}}

\caption{The experimental setup studied for possible 
parity-violation measurement.\label{fig:exp}}

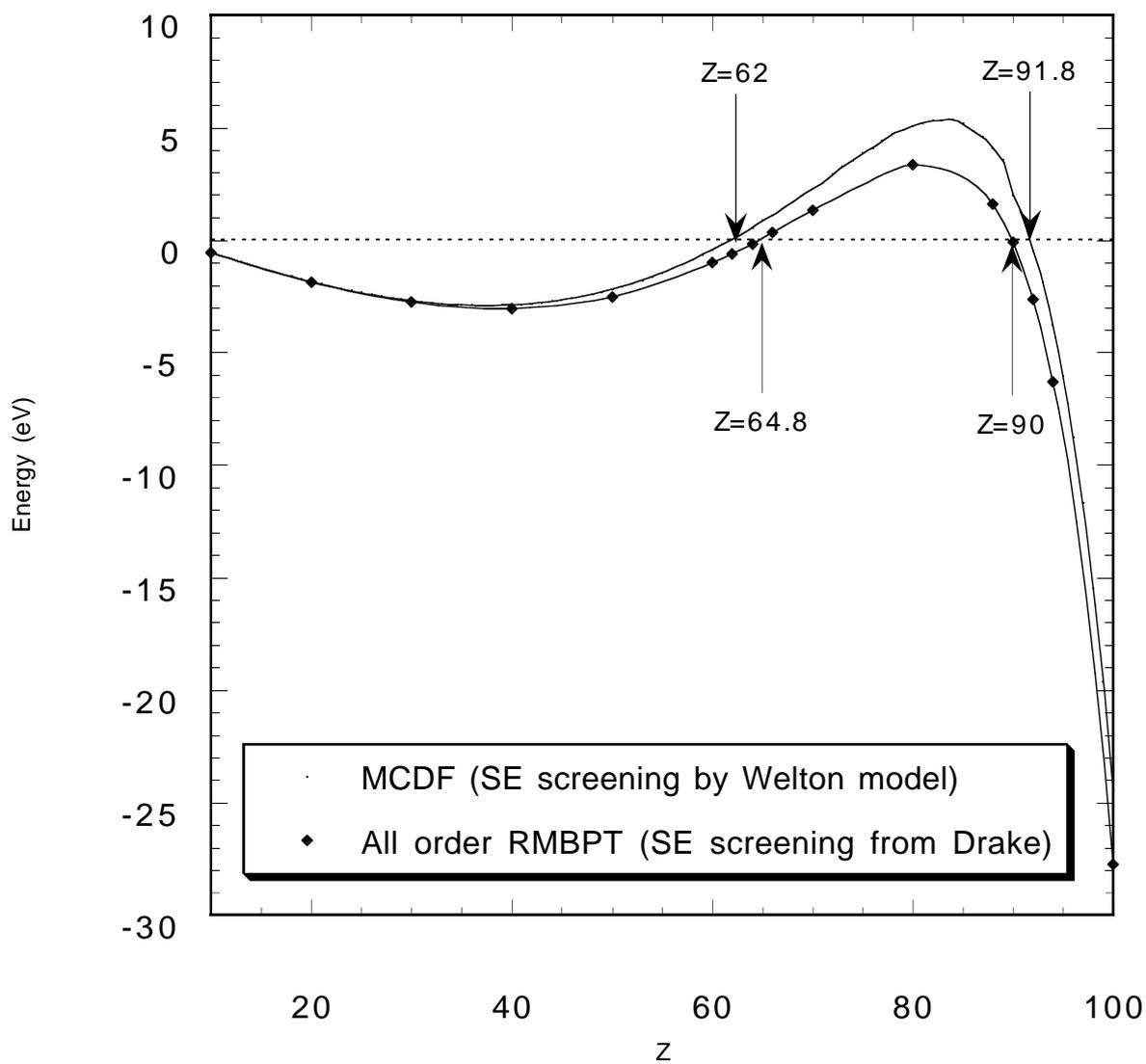
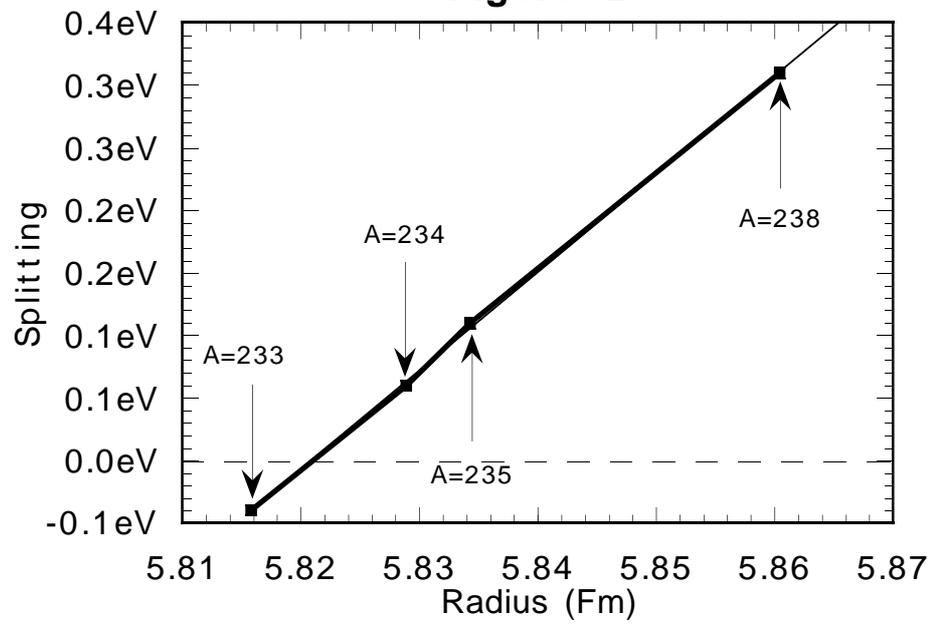
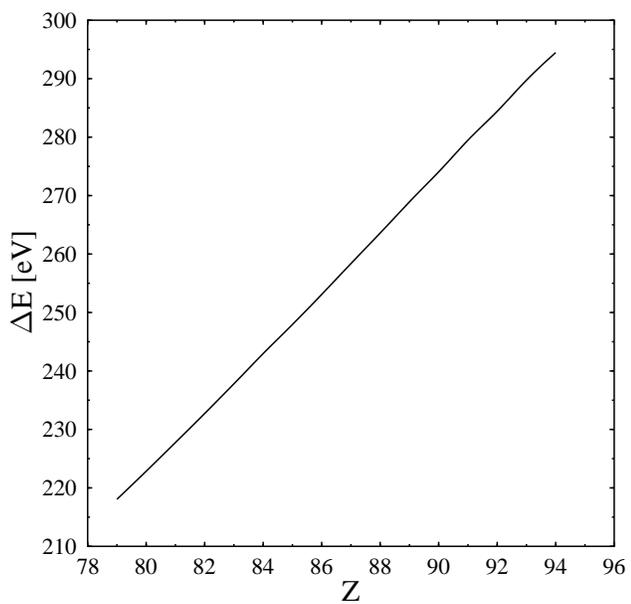
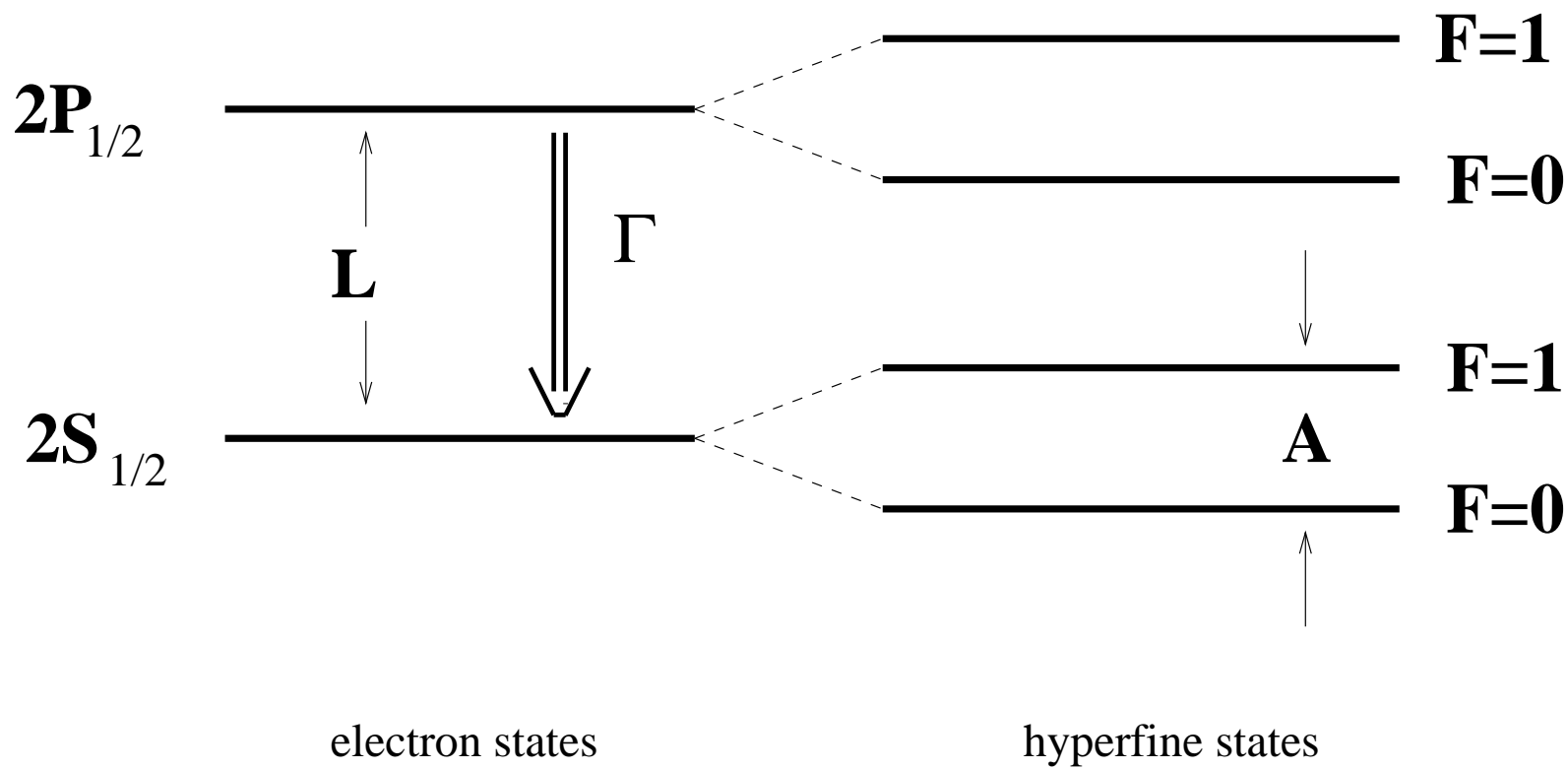
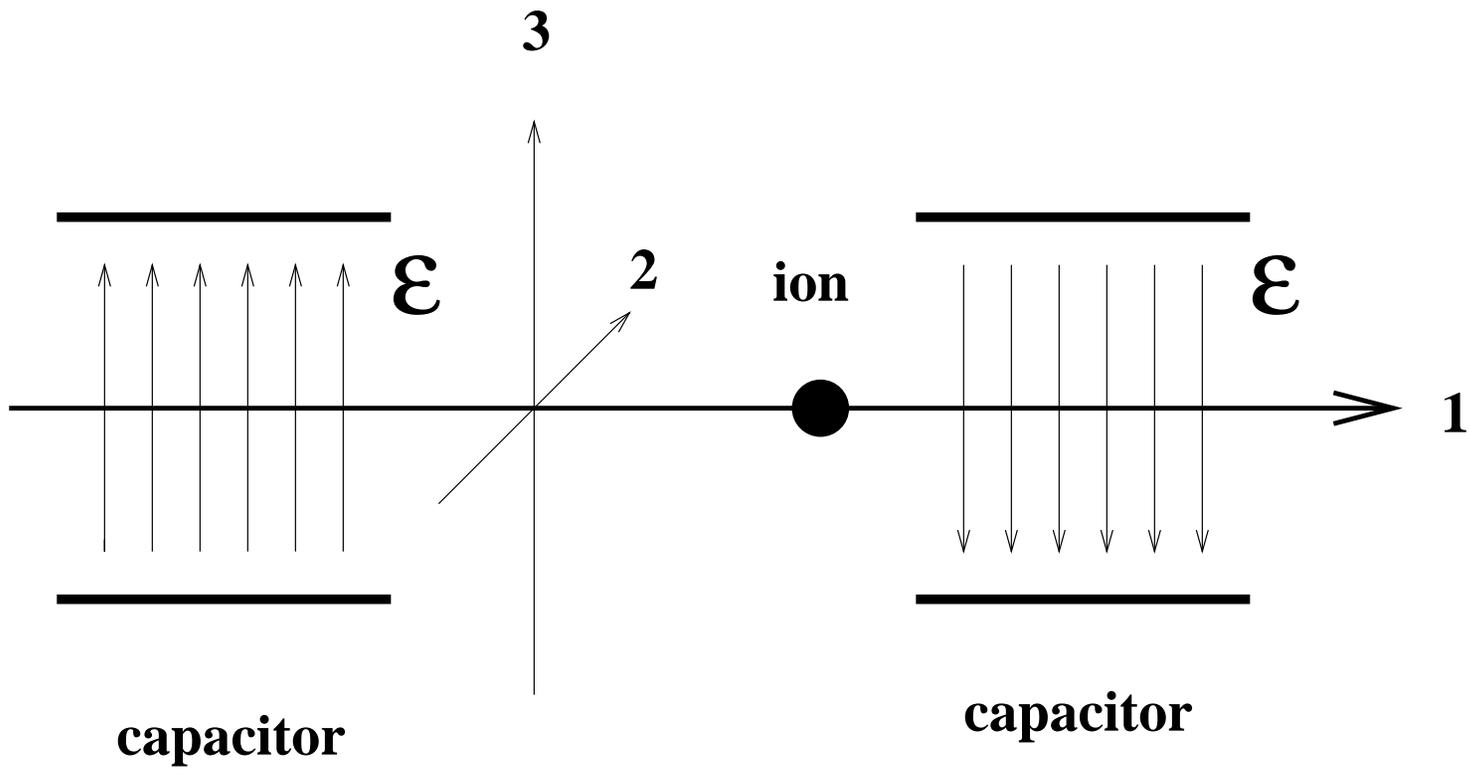
\end{document}